\documentstyle[prc,aps,preprint,epsfig]{revtex}

\begin{document}
\draft
\date{\today}

\title{Optimization of finite-range effective interaction for 
in-medium cross sections}

\author{Denis Lacroix$^{a)}$, Sakir Ayik$^{b)}$, Osman Yilmaz$^{c)}$, 
and Ahmet Gokalp$^{c)}$}
\address{$^{a)}$ {\it 
LPC/ISMRA, Blvd du Mar\'{e}chal Juin, 14050\ Caen, France,} \\
$^{b)}$ {\it Tennessee Technological University, Cookeville TN38505, USA}\\
$^{c)}$ {\it Physics department, Middle East Technical University, 06531
Ankara, Turkey.} \\
\begin{abstract} 
In order to incorporate the finite range effect into effective 
interactions, a modification of the Skyrme force by 
introducing a cut-off factor for high momentum transfers is
proposed. The parameters of the cut-off factor are determined by
fitting the microscopic in-medium cross-sections of Li and Machleidt
over a wide range of energy and nuclear density. Results for 
the SkM$^*$ and SLy4  forces are presented.
\end{abstract}
}

\maketitle 

{\bf PACS: 3.75.Cs, 21.30.-x, 25.70.-z} 



\section{Introduction}

Effective interactions, like Skyrme \cite{Sky56,Vau72} 
or Gogny \cite{Gog75} forces, 
associated with the mean-field theory have made 
it possible to describe a large body of nuclear
properties along the whole nuclear chart \cite{Rin81}. 
These forces are constantly refined to reproduce
more and more precisely experimental observations 
\cite{Cha98}. However, these  effective forces encounter 
difficulties for describing quantities that are 
sensitive to the short range correlations such as
nucleon-nucleon collisions cross-sections 
\cite{Cas90,Yil00}. For example, the nucleon-nucleon 
cross-sections calculated from Skyrme forces do not 
interpolate correctly between free-space and 
the medium, and reaches large values at low densities
due to the 
unrestricted high momentum transfers. These difficulties 
become more dramatic in the applications of the 
extend mean-field approach to nuclear
dynamics \cite{Ayi00-1}. Recently, 
it has been pointed out that serious difficulties 
arise when an effective Skyrme force is employed in the 
collision term of the extended Time-Dependent 
Hartree-Fock theory \cite{Lac00}. More generally, 
such shortcomings are expected in transport theories 
that go beyond the mean-field approximation by including
two-body correlations through residual interactions. 
In particular, in the applications of Boltzmann-type 
semi-classical transport models \cite{Ber88,Ayi88,Cho96} 
or molecular dynamics calculations \cite{QMD},
the effective interactions are rarely adjusted for 
in-medium cross-sections, which may strongly influence 
the collision dynamics. In this work, we discuss an 
approximate treatment of incorporating the effect of
finite range into Skyrme parameterization of the effective
interactions, and optimize the parameters by fitting the
nucleon-nucleon cross-sections to the microscopic in-medium 
cross-sections of Li and Machleidt \cite{Li93} 

\section{Effective forces with finite range.}

Following ref. \cite{Ayi00-1,Cho00,Lac00}, we propose 
a modification of Skyrme interactions by 
introducing a cut-off factor
${\cal C} \left(|<{\bf q}^{2}>|\right)$  in the two-body 
interaction matrix elements of the form,
\begin{equation}
<ij|v|kl>=<ij|v_{S}|kl>\cdot {\cal C} 
\left(|<ij|{\bf q}^{2}|kl>|\right)   
\label{r_gen}
\end{equation}
where $v_{S}$ represents $t_0$ and $t_3$ terms of the Skyrme 
interaction,
\begin{eqnarray}
v_{S} = t_0(1+x_0P_\sigma )\delta \left( \mathbf{r}\right) 
+ \frac 16t_3(1+x_3P_\sigma )\left[ \rho \left( \mathbf{R}\right)
\right] ^\alpha \delta \left( \mathbf{r}\right)
\end{eqnarray}
with ${\bf r}={\bf r}_{1}-{\bf r}_{2}$ and $\mathbf{R} = 
\left( {\bf r}_{1}+{\bf r}_{2}\right)/2$.
We assume that the cut-off factor ${\cal C}$ is a function of
a matrix $<ij|{\bf q}^{2}|kl>$, which is determined by the
expression, 
\begin{eqnarray}
<ij|\delta ({\bf r})|kl><ij|{\bf q}^{2}|kl>=
<ij|{\bf q}^{2}\delta ({\bf r})+\delta ({\bf r}){\bf q}^{2}|kl> 
\end{eqnarray}
where
${\bf q}=({\bf p}_{1}-{\bf p}_{2})/2$ represents the relative
momentum operator. This quantity 
provides a measure for the relative momentum in finite systems,
and for the special case of nuclear matter,
the function ${\cal C}$ represents a cut-off in momentum space.
The expression (\ref{r_gen}) is very convenient for practical 
applications, not only for nuclear matter but also for finite 
systems, since it does not introduce additional numerical
effort as compared to Skyrme forces. 

We can calculate the nucleon-nucleon
cross-sections associated with the modified effective force 
(\ref{r_gen}). In the local density approximation, the expression 
of the cross-section is given by,
\begin{eqnarray}
\left(\frac{d \sigma}{d \Omega} \right)_{xy} &=& \frac{1}{4 \pi}
{\cal R}_{xy} \left(\rho\right) 
\left[\overline{{\cal C}}\left(|<{\bf q}^{2}>|\right)\right]^2 
\label{cross}
\end{eqnarray}
where ${\cal R}_{xy} \left(\rho\right)$  account for the 
$t_0$ and $t_3$ part and involves density dependence, 
while $\overline{{\cal C}}$ comes from the cut-off
factor. In this expression, the labels $xy$ denote the 
cross-section in the proton-neutron channel ($np$), 
the proton-proton channel ($pp$) and the neutron-neutron 
channel ($nn$), which are given by\cite{Ayi98},
\begin{eqnarray}
\begin{array} {cl}
{\cal R}_{pn} \left(\rho\right) =\frac{\pi m^2}
{4 \hbar \left(2\pi \hbar\right)^3}
\frac{1}{2} & \left( \left[t_0(1-x_0) +
\frac{t_3}{6}(1-x_3)\rho^{\alpha}\right]^2 \right. \\
& \left. + 3\left[t_0(1+x_0) + \frac{t_3}{6}(1+x_3)\rho^{\alpha}\right]^2
\right)
\end{array}
\end{eqnarray}
and 
\begin{eqnarray}
\begin{array} {cl}
{\cal R}_{pp} \left(\rho\right) =\frac{\pi m^2}{4 \hbar
\left(2\pi \hbar\right)^3}
& \left[t_0(1-x_0) +
\frac{t_3}{6}(1-x_3)\rho^{\alpha}\right]^2 
\end{array}
\end{eqnarray}
with a similar expression for the neutron-neutron cross-section. 
The spin-isospin averaged total cross-sections is given by 
$\sigma_{tot} = \left(\sigma_{nn} + \sigma_{pp} + 2 \sigma_{pn}\right)/4$,
where $\sigma_{xy} =2\pi 
\int sin \Theta d\Theta \left( d\sigma/d\Omega\right)_{xy}$ denotes the
total cross-section in the corresponding channel. In the following,
we refer to the density-dependent term associated with the total 
cross-section as ${\cal R}_{tot}$.

The choice of the cut-off ${\cal C}$ factor is not unique. 
For simplicity, we may parameterize the cut-off in terms of 
Gaussian or exponential functions.  These cut-off functions, 
respectively, lead to a Gaussian and Breit-Wigner shapes  
in the momentum space,
\begin{eqnarray}
\overline{{\cal C}} (q,q')&=& e^{-\beta^2 (q^2 + q'^2)/\hbar^2} \\
&& \\ \nonumber
\overline{{\cal C}} (q,q')&=& \frac{1}{1 + \beta^2 (q^2 + q'^2)/\hbar^2}
\end{eqnarray}
where ${\bf q} = ({\bf p}_1-{\bf p}_2)/2$ and 
${\bf q}'=({\bf p}_3-{\bf p}_4)/2$ represent the relative momenta before
and after collisions. For elastic collisions, we have $q=q'$, thus
the energy is given by $E_{lab} = 2q^2/m$. 
\begin{figure}[tbph]
\begin{center}
\epsfig{file=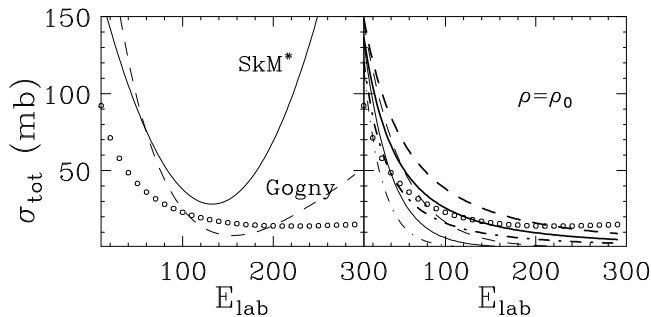,width=9.cm}
\end{center}
\caption{Spin-isospin averaged cross-section as function
of energy at the normal nuclear density. Left panel: the result
obtained from the SkM$^*$ force(solid line) and Gogny force 
(dashed line) are compared with the microscopic cross-sections of
Li and Machleidt (circles). Right panel: total cross-sections 
calculated for the SkM$^*$ force with bare nucleon mass,
using a Gaussian cut off (thick line) and a Breit-Wigner cut-off
(thick line). In both cases, the results with $\beta=0.5$ 
fm (dashed line), $\beta=0.7$ fm (solid line) and
$1$ fm (dot-dashed line) are presented. }
\label{fig:cut1}
\end{figure}

Figure \ref{fig:cut1} shows a typical example 
of energy dependence of in-medium cross-sections at normal 
nuclear matter density $\rho_0$. The results on the right panel
are obtained with the SkM$^{*}$ force \cite{Bar82} and using 
a Gaussian (thin lines) or a Breit-Wigner
(thick lines) cut-off for different values of $\beta$. 
These results are compared to the microscopic calculations of
Li and Machleidt, which are shown by circles. Left panel of 
figure \ref{fig:cut1} presents the energy dependence of cross-sections
that are obtained using the full Skyrme SkM$^{*}$ force and 
the Gogny force\cite{Gog75}.
From this figure, we first remark that 
the standard effective forces, like Skyrme or Gogny forces, 
provide a poor approximation for the energy dependence of the
nucleon-nucleon cross-sections. We, also, see 
that calculations with the Gaussian or the Breit-Wigner cut-off 
are considerably improved and provide
an approximate description for the cross-sections of Li and 
Machleidt \cite{Li93}. However, a cut-off factor with  
a single $\beta$ value is not able to
reproduce the behavior of the cross section for a wide-range 
of energy. For example, with a Gaussian factor, $\beta=1$ fm 
gives a good approximation for the low energy part, 
while $\beta=0.7$ fm is better for the high energy part. 

In order to improve the description
of the cross-section in the whole energy range, 
we propose the following cut-off factor,
\begin{eqnarray}
\overline{{\cal C}} (q,q') = \frac{1 + \beta_1^2 (q^2 + q'^2)/\hbar^2}
                      {1 + \beta_2^2 (q^2 + q'^2)/\hbar^2}
\end{eqnarray} 
\begin{table}[tbp]
\begin{center}
\begin{tabular}{ccccc}
Density & & ${\cal R}_{tot}(\rho)$  & $\beta_1$  & $\beta_2$  \\
\hline
$\rho_0$   & {\it pp}  & 243.2 & 1.15 & 5.15 \\ 
           & {\it np}  & 209.7 & 0.28 & 1.49 \\
           & {\it tot} & 226.5 & 0.60  & 2.80 \\
\hline
$\rho_0/2$ & {\it pp}  & 212.6 & 0.52 & 2.38 \\  
           & {\it np}  & 417.9 & 0.31 & 1.87 \\
           & {\it tot} & 315.3 & 0.37 & 2.03 \\
\hline
$\rho_0/3$ & {\it pp}  & 197.2 & 0.36 & 1.64 \\  
           & {\it np}  & 585.4 & 0.38 & 2.29 \\
           & {\it tot} & 391.3 & 0.38 & 2.09 \\
\end{tabular}
\end{center}
\caption{ Parameters $\beta_1$ and $\beta_2$ associated with the
SLy4 force, which are determined by fitting the
microscopic cross-sections of Li and Machleidt at different densities.
We also present the ${\cal R}_{tot}(\rho)$ values for 
the same force at different densities. }
\label{tab:cut_pnn_sly4}
\end{table}

\begin{table}[tbp]
\begin{center}
\begin{tabular}{ccccc}
Density & & ${\cal R}_{tot}(\rho)$  & $\beta_1$  & $\beta_2$  \\
\hline
$\rho_0$   & {\it pp}  & 47.8  & 0.05 & 0.28 \\ 
           & {\it np}  & 324.9 & 0.61 & 3.18 \\
           & {\it tot} & 186.4 & 0.44 & 2.03 \\
\hline
$\rho_0/2$ & {\it pp}  & 103.2 & 0.12 & 0.63 \\  
           & {\it np}  & 506.3 & 0.45 & 2.60 \\
           & {\it tot} & 304.8 & 0.35 & 1.92 \\
\hline
$\rho_0/3$ & {\it pp}  & 141.4 & 0.18 & 0.88 \\  
           & {\it np}  & 620.4 & 0.42 & 2.52 \\
           & {\it tot} & 380.9 & 0.35 & 1.91 \\
\end{tabular}
\end{center}
\caption{Same as table \protect{\ref{tab:cut_pnn_sly4}} for the SkM$^{*}$
force.}  
\label{tab:cut_pnn_skms}
\end{table}

By comparing the calculations with this cut-off factor with the
microscopic  ({\it pp}) and ({\it np}) cross-sections, we 
deduce the best corresponding $\beta_1$ and $\beta_2$ coefficients 
for each channel and for different nuclear densities. The extracted
values of $\beta_1$ and $\beta_2$, that are associated with 
SLy4 \cite{Cha98} and SkM$^{*}$ forces, are reported 
in table \ref{tab:cut_pnn_sly4} and
table \ref{tab:cut_pnn_skms}, respectively. 
We observe from these tables that for diluted systems the coefficients
are less sensitive to nuclear density. Here, we consider the spin-isospin
averaged cross-sections. For this purpose, by averaging over spin-isospin 
channel and also over different densities, we deduce average values of 
$\beta_1=0.48$ fm and $\beta_2=2.41$ fm for the SLy4 force, and 
$ \beta_1=0.29$ fm and $\beta_2 =1.66$ fm for the SkM$^*$ force. 
In table \ref{tab:force}, we present different parameters 
for modified Skyrme forces, which are indicated by  
(SLy4)$^{cut}$ and {(SkM$^*$)}$^{cut}$.
\begin{figure}[tbph]
\begin{center}
\epsfig{file=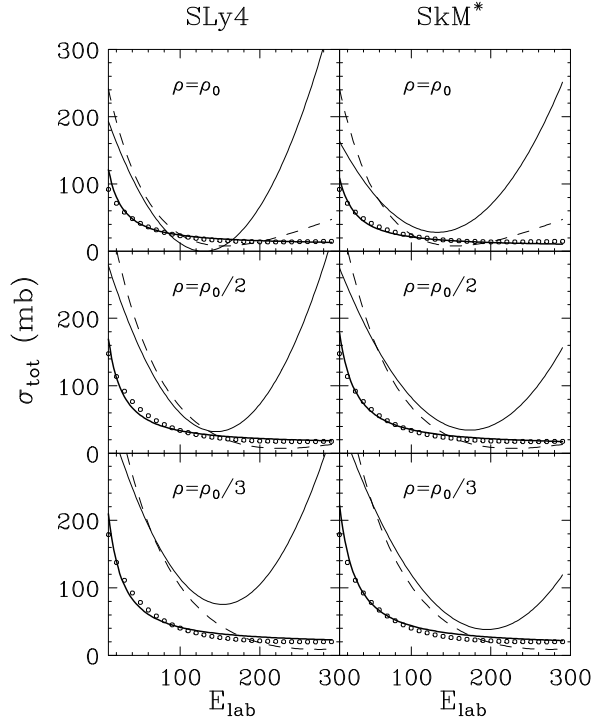,width=8.cm}
\end{center}
\caption{Spin-isospin averaged cross-section
as a function of energy at different nuclear densities.
Left panel: thick lines show the result from the modified
force (SLy4)$^{cut}$
with the new cut-off factor with 
$\beta_1=0.39$ fm and 
$\beta_2=2.05$ fm and a comparison with the cross-sections
of Li and Machleidt indicated by circles.
Right panel: same as left panel for the modified force 
SkM$^{*}$ force with $\beta_1=0.29$ fm and 
$\beta_2 =1.66$ fm. In figure, thin and dashed lines show
the results of the full SLy4 Skyrme force and the Gogny force,
respectively.}
\label{fig:cut2}
\end{figure}

\begin{figure}[tbph]
\begin{center}
\epsfig{file=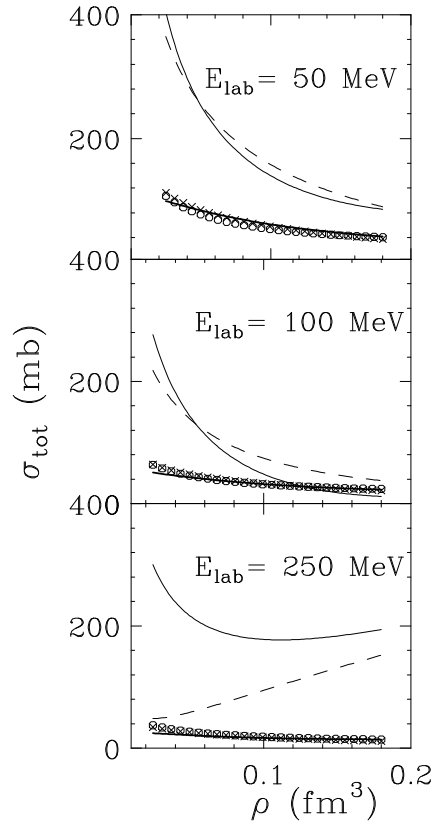,width=6.cm}
\end{center}
\caption{Spin-isospin averaged cross-sections as a function of density 
at different energies. The results for the (SkM$^{*}$)$^{cut}$ force,
the (SLy4$^{*}$)$^{cut}$ force and cross-sections of Li and Machleidt 
are shown by crosses, circles and solid lines, respectively.
Dashed lines and thin lines
lines represents the results obtained with the full Skyrme forces, 
Sly4 and SkM$^{*}$, respectively.}
\label{fig:elab}
\end{figure}

Figure \ref{fig:cut2} shows the total nucleon-nucleon cross-sections
as a function of bombarding energy at  nuclear densities
$\rho=\rho_0$, $\rho=\rho_0/2$ and $\rho=\rho_0/3$. 
The thick lines show the result of our calculations, left panel
for (SLy4)$^{cut}$ and right panel for {(SkM$^*$)}$^{cut}$, 
with the parameters
reported in table \ref{tab:force}. As seen from the figure, our calculations
perfectly match the in-medium cross-sections of Li and Machleidt, which
are indicated by circles. Figure \ref{fig:elab} presents the total 
cross-section
as a function of nuclear density at energies $E_{lab}=50, 100, 250 MeV$.
Again, a perfect agreement with the microscopic
cross-sections is found for a wide range of nuclear density.
In figures \ref{fig:cut2} and \ref{fig:elab}, dashed lines and thin 
lines show 
the results of the Gogny force and the full SLy4 force, respectively.
\begin{table}[tbp]
\begin{center}
\begin{tabular}{c|ccccccc}
 & $t_0$  & $t_3$ & $x_0$ & $x_3$& 
$\alpha $& $\beta_1$  &  $\beta_2$ \\
 & (MeV.fm$^3$) & (MeV.fm$^{5/2}$) &
& & & (fm) & (fm) \\
\hline
(SLy4)$^{cut}$   &-2488.91& 13777 & 0.834 &  1.354 & $\frac 16$  & 0.39 & 2.05\\
{(SkM$^*$)}$^{cut}$&-2645   & 15595 & 0.09  &  0.0   & $\frac 16$ & 0.29 & 
1.66\\
\end{tabular}
\end{center}
\caption{ Summary of parameters for the modified Skyrme forces 
(SLy4)$^{cut}$ and {(SkM$^*$)}$^{cut}$.} 

\label{tab:force}
\end{table}

\section{Conclusion}

In this paper, in order to provide an approximate
description of the in-medium nucleon-nucleon cross-sections,
we discuss a modification of the effective 
Skyrme interactions by introducing a cut-off factor for
large momentum transfer.
We show that neither Gaussian nor Breit-Wigner cut-off in 
momentum space are appropriate for this purpose.  
We propose a new parameterization of the cut-off function 
and determine its parameters
by fitting the microscopic in-medium cross-sections of Li and Machleidt,
which interpolate correctly between the free-space and the medium
and provide the best available in-medium cross-sections.
The proposed modification approximately takes account for the finite range
effect of the residual interactions, and provides a practical but realistic
tool for application of transport models, in which the short range 
correlations are incorporated in the form of a binary collision term.

{\bf Acknowledgments}
      
We thank Ph. Chomaz for discussion at the initial stage of this work
One of us (S. A.) gratefully acknowledges GANIL Laboratory for a partial
support and warm hospitality extended to him during his visit to Caen. This
work is supported in part by the US DOE grant No. DE-FG05-89ER40530.

\end{document}